\documentclass[aps,prl,twocolumn,showpacs]{revtex4}
\usepackage{amssymb,amsmath}
\usepackage{graphicx}
\usepackage{dcolumn}
\usepackage{bm}
\usepackage[colorlinks=true,dvipdfm]{hyperref}

\begin{document}

\title{Experimental Demonstration of the Unified Framework for the Mixed
State Geometric Phase}
\author{Jiangfeng Du$^{1,2}$}
\email{djf@ustc.edu.cn}
\author{Mingjun Shi$^{1}$}
\author{Jing Zhu$^{1}$}
\author{Vlatko Vedral$^{3,4}$}
\author{Xinhua Peng$^{2}$}
\author{Dieter Suter$^{2}$}
\affiliation{$^1$ Hefei National Laboratory for Physical Sciences at Microscale and
Department of Modern Physics, University of Science and Technology of China,
Hefei, Anhui 230026, People's Republic of China \\
$^2$Fachbereich Physik, Universit\"{a}t Dortmund, 44221 Dortmund, Germany $%
^3 $Department of Physics and Astronomy, University of Leeds, Leeds LS2 9JT,
UK\\
$^{4}$Department of Physics, National University of Singapore, 2 Science
Drive, Singapore}

\begin{abstract}
Geometrical phases have been applied in virtually every major branch of physics and they play an important role in topology and knot theory in mathematics and quantum computation. However, most of the early works focus on pure quantum states which are very unrealistic and almost never occur in practice. The two existed definitions -Uhlmann's and Sj\"{o}qvist's- would result in different values for the geometric phase in general. The definition of geometric phase in mixed state scenario is still an open question. Here we present a unified framework for both approaches within one and the same formalism based on simple interferometry. We then present experimental results which confirm both approaches to mixed state geometric phase and clearly demonstrate that their unification is possible. Our experiments are furthermore the first such to measure Uhlmann's mixed state geometric phase and show in addition that it is different to the Sj\"{o}qvist's phase.
\end{abstract}

\maketitle

When a quantum system undergoes a general physical evolution, its state gains two phase factors, the dynamical and the geometric. The latter is independent of the former and its value is a function only of the path described by the system throughout its evolution; in particular it does not depend on any details of the dynamics\cite{1,2}. Geometric phases play an extremely important role in many different areas of physic. Originally discovered within the context of quantum mechanics and atom optics, their significance quickly became transparent in mathematical physics, condensed matter and high energy physics\cite{3}. Now, they have even been used in quantum computation\cite{4,5,6,7}.

Most of the early works on the geometrical phase focus on pure quantum states\cite{8,9,10,11}. These, however, are very unrealistic and almost never occur in practice. The difficulty with mixed states is their reduced coherence, which makes any notion of a phase more difficult to define and measure. Two main definitions of mixed state geometric phase exist\cite{12,13}. One is based on the fact that any mixed state can be represented as a pure one if we allow ancillas. We then have to optimise over many purification and this was done by Uhlmann\cite{12}. The other way of defining mixed state geometric phase, due to Sj\"{o}qvist et al\cite{13}, is operational\cite{14,15}: it uses the same interferometric procedure that reproduces pure state geometric phase. These two definitions result in different values for the geometric phase in general. Here we present a unified framework for both approaches within one and the same formalism based on simple interferometry. We then present experimental results within liquid-state NMR which confirm both approaches to mixed state geometric phase and clearly demonstrate that their unification is possible. Our experiments are furthermore the first such to measure Uhlmann's mixed state geometric phase and show in addition that it is different to the Sj\"{o}qvist et al phase.

To describe Sj\"{o}qvist et al's and Uhlmann's definitions of the mixed state geometric phase within a unified picture, we consider a mixed state of a spin half nucleus which evolves with time in a uniform external magnetic field, adopting the nomenclature from quantum information science we will sometimes refer to spin half particles as qubits.

The initial state of the qubit is $\rho _{s}=p_{1}|0\rangle \langle0|+p_{2}|1\rangle \langle 1|$. Here $p_{1}$ and $p_{2}$ are the eigenvalues of $\rho _{s}$ and represent the probabilities for the qubit to be in states $|0\rangle $ and $|1\rangle $ respectively. The states $|0\rangle $ and $|1\rangle $ are the corresponding eigenstates and also the eigenstates of the Pauli matrix $\sigma _{z}$. The magnetic field is $\mathbf{B}_{s}=\mathbf{e}_{x}B_{s}\sin \theta _{s}+\mathbf{e}_{z}B_{s}\cos \theta _{s}$. Then the time evolution of the system qubit is determined by the unitary operator $U_{s}(t)=exp[-iH_{s}t/\hbar ]$, where $H_{s}=-\mathbf{\mu }\cdot \mathbf{B}_{s}$ is the Hamiltonian of the system and $\mathbf{\mu }$ is the magnetic moment of system qubit. With the help of an ancillary qubit, the purification of $\rho _{s}$ can be expressed as $|\Psi \rangle =\sqrt{p_{1}}|0\rangle _{s}|0\rangle _{a}+\sqrt{p_{2}}|1\rangle _{s}|1\rangle _{a}$, where the subscript $s$ indicates the system qubit and $a$ the ancillary qubit. Although the evolution of the ancilla does not (and cannot) affect the form of $\rho _{s}$, it does affect the phase of the pure components of $\rho _{s}$.

We now explain how Sj\"{o}qvist's and Ulmann's mixed geometric phase emerge from two different purification evolutions. That is, by means of controlling the evolution of the ancilla, we can obtain different geometric phases of the mixed system state.

\textit{Case 1 Sj\"{o}qvist phase} In this case we let the unitary evolution of the ancilla, $U_{a}^{Sjo}(t)$, be determined by the magnetic field. As a result, the evolution of the ancilla cancels the dynamical phase of $|0\rangle _{s}$ and $|1\rangle _{s}$ under the influence of $U_{s}(t)$. Hence the purification state $|\Psi \rangle $ is parallel transported. The corresponding geometric phase is $arg[\langle \Psi |U_{s}(t)\otimes U_{a}^{Sjo}(t)|\Psi \rangle ]$, which is just Sj\"{o}qvist's mixed state geometric phase 
\begin{equation}
arg[p_{1s}\langle 0|U_{a}^{Sjo}(t)|0\rangle _{s}+p_{2s}\langle
1|U_{a}^{Sjo}(t)|1\rangle _{s}].  \notag
\end{equation}

\textit{Case 2 Uhlmann phase} The other non-trivial option for the evolution of ancilla, denoted as $U_{a}^{Uhl}(t)$, is associated with the magnetic field $\mathbf{B}_{a}=-\mathbf{e}_{x}B_{a}\sin \theta _{a}-\mathbf{e}_{z}B_{a}\cos \theta _{a}$ with two conditions: (a) $\tan \theta _{a}=2\sqrt{p_{1}p_{2}}\tan \theta _{s}$ and (b) $B_{a}\cos \theta _{a}=B_{s}\cos \theta_{s}$. Then, under the transformation $U_{s}(t)\otimes U_{a}^{Uhl}(t)$, $|\Psi \rangle $ is parallel transported and we acquire the corresponding geometric phase $arg[\langle \Psi |U_{s}(t)\otimes U_{a}^{Uhl}(t)|\Psi\rangle ]$. It can be proved that this geometric phase is just Uhlmann's phase, which is given by 
\begin{widetext}
\begin{equation}
-\arctan[\frac{(p_1-p_2)(\cos\theta_s\tan\frac{\omega_s t}{2}+\cos\theta_a\tan\frac{\omega_a t}{2})}{1+(\cos\theta_s\cos\theta_a-\sqrt{p_1 p_2}\sin\theta_s\sin\theta_a)\tan\frac{\omega_s t}{2}\tan\frac{\omega_a t}{2}}],\nonumber
\end{equation}
\end{widetext}where $\omega _{s}=\mu B_{s}$ is the procession frequency of system qubit in magnetic field $\mathbf{B}_{s}$ and similarly for $\omega_{a}$, and without loss of generality we let $p_{1}>p_{2}$.

\begin{figure}[tbph]
\centering
\includegraphics[width=1\columnwidth]{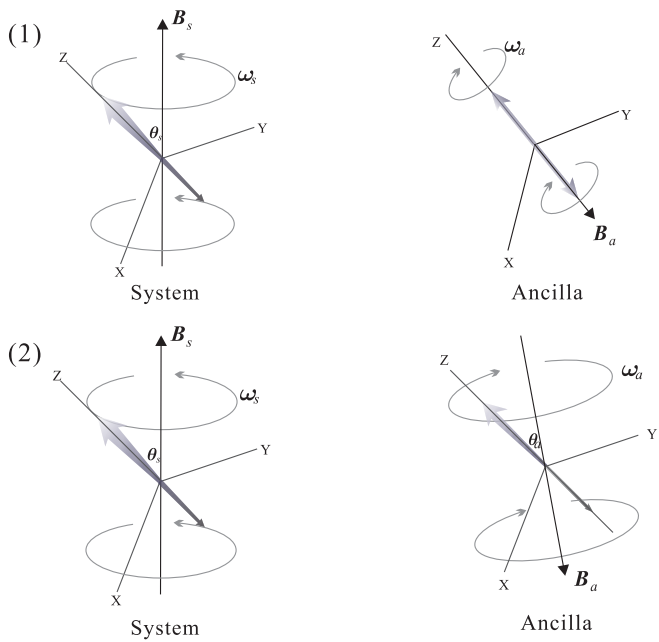}
\caption{(Color online) Unified picture of mixed state geometric. Two pure components of the system's mixed state are represented by two arrows along the $z$ axis. The fixed magnetic field $\mathbf{B}_{s}$ lies in the xz-plane, at an angle $\protect\theta _{s}$ from the z axis. The system qubit rotates around $\mathbf{B}_{s}$ with angular frequency $\protect\omega_{s}=\protect\mu B_{s}$. The upper figure (1) corresponds to case 1, in which the magnetic field for the ancilla is $\mathbf{B}_{a}=-\mathbf{e}_{z}B_{s}\cos \protect\theta _{s}$. The lower figure (2) corresponds to case2. Here the ancilla evolves in the magnetic field $\mathbf{B}_{a}=-\mathbf{e}_{x}B_{a}\sin \protect\theta _{a}-\mathbf{e}_{z}B_{a}\cos \protect\theta _{a}$, where the parameters satisfy two conditions: $\tan \protect\theta _{a}=2\protect\sqrt{p_{1}p_{2}}\tan \protect\theta _{s}$ and $B_{a}\cos \protect\theta _{a}=B_{s}\cos \protect\theta _{s}$.}
\label{eps1}
\end{figure}
\begin{figure}[tbph]
\centering
\includegraphics[width=0.8\columnwidth]{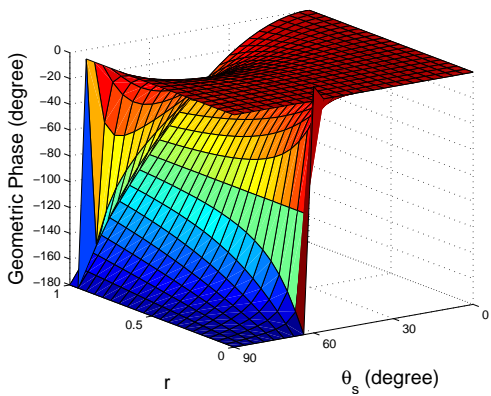}
\caption{(Color online) The two cyclic geometric phases, the Uhlmann geometric phase (upper layer) and the Sj\"{o}qvist geometric phase (lower layer), as a function of purity $r$ and angle $\protect\theta _{s}$. The procession frequency of system qubit $\protect\omega _{s}$ is fixed by $\protect\omega _{s}t=\protect\mu B_{s}t=2\protect\pi $ so that the system qubit undergoes a cyclic evolution. }
\label{eps2}
\end{figure}

The evolution paths of the system and the ancilla are illustrated in Fig. \ref{eps1}. In Fig. \ref{eps2}, we plot Sj\"{o}qvist's and Uhlmann's cyclic phase as functions of the angle $\theta _{s}$ and the purity $r$ ($r=p_{2}-p_{1}$) of the mixed state of the system. 

\begin{figure}[tbph]
\centering
\includegraphics[width=1\columnwidth]{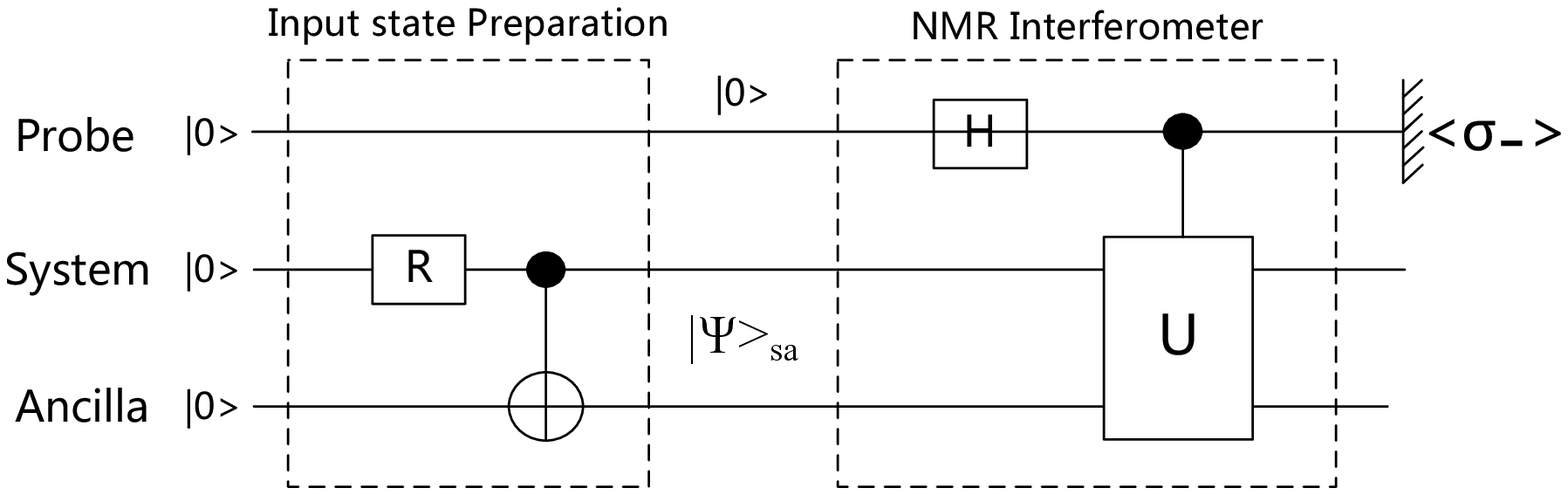}
\caption{The quantum network for observing geometric phase using NMR interferometry. Qubit-1 is the probe qubit, which is used to observe the geometric phase. The pseudo-Hadamard gate $H=e^{-i\frac{\protect\pi }{2}\protect\sigma _{y}}$ puts the probe qubit into an equal superposition state, acting as a symmetric beam splitter. Another single-qubit rotation $R=e^{-i\arccos (\protect\sqrt{(}p_{1}))\protect\sigma _{y}}$ and a controlled-NOT gate prepare qubits 2 and 3 into an entangled state $|\Psi_{in}\rangle _{sa}$. represents a bilocal unitary for either the Uhlmann phase or the Sj\"{o}qvist phase (see the text for details), performed on the input state $|\Psi _{in}\rangle _{sa}$ conditionally on the state of the probe qubit. }
\label{eps3}
\end{figure}

Having presented the unified picture of two phases of the mixed state, we can also observe them in a unified experimental setting. Usually any phase variation is observed by some kind of an interferometer\cite{8}. Here we resort to NMR interferometry and the schematic diagram of the experiment is presented in Fig. \ref{eps3}. The system qubit is initially in a mixed state, which is purified to the state $|\Psi _{in}\rangle _{sa}$ by using an ancilla qubit (see the first part of Fig. \ref{eps3}).

The goal of the experiment is to observe the geometric phase pertaining to the system qubit which is acquired when this qubit experiences a unitary evolution. For this purpose, we have to introduce another qubit as the probe qubit that is coupled to the system and ancilla. As shown in Fig. \ref{eps3}, the probe qubit is initialized into an equal superposition $(|0\rangle_{p}+|1\rangle _{p})/\sqrt{2}$ by applying a Hadamard gate to the $|0\rangle_{p}$ state. Then a conditional rotation only brings a unitary evolution $U$ on the \textquotedblleft copy\textquotedblright\ of the $|\Psi _{in}\rangle_{sa}$ state that is connected to the $|1\rangle _{p}$ state. Here $U$ is a bilocal unitary evolution and can be either $U_{s}(t)\otimes U_{a}^{Sjo}(t)$ for the Sj\"{o}qvist phase (Case 1) or $U_{s}(t)\otimes U_{a}^{Uhl}(t)$ for Uhlmann's phase (Case 2). After this, the system reaches the state $(|0\rangle _{1}\otimes \Psi _{in}\rangle _{23}+|1\rangle _{1}\otimes U\Psi_{in}\rangle _{23})/\sqrt{2}$, so that the probe qubit is in the state $[I+Re_{sa}\langle \Psi _{in}|U|\Psi _{in}\rangle _{sa}\sigma_{x}^{p}+Im_{sa}\langle \Psi _{in}|U|\Psi _{in}\rangle _{sa}\sigma_{y}^{p}]/2$. By measuring the expectation value $\langle \sigma_{-}^{p}\rangle =\langle \sigma_{x}^{p}-i\sigma_{y}^{p}\rangle $ of the probe qubit, we obtain the geometric phase, $arg(\langle \Psi _{in}|U|\Psi_{in}\rangle )=arg(\langle \sigma _{-}^{p}\rangle )$. 

The molecule ($^{13}C$-labelled alanine) used for this experiment contains three $^{13}C$ spin-$\frac{1}{2}$ nuclei as qubits. The Hamiltonian of the 3-qubit system is (in angular frequency units) $H =\sum_{i=1}^{3} \omega_{i}I_{z}^{i}+2\pi\sum_{i<j}^{3}J_{ij}I_{z}^{i}I_{z}^{j}$ with the Larmor angular frequencies of the $i^{th}$ spin $\omega_i$ and spin-spin coupling constants $J_{13}=54.1$Hz, $J_{23}=34.9$Hz, $J_{12}=-1.3$Hz. Qubits 1, 2 and 3 are used as the probe qubit, system qubit, and ancillary qubit respectively. Experiments were performed at room temperature using a
standard 400MHz NMR spectrometer (AV-400 Bruker instrument).

The system was first prepared in a pseudo-pure state(PPS) $\rho _{000}=\frac{1-\epsilon }{8}\mathbf{1}+\epsilon |000\rangle \langle 000|$, where $\epsilon \approx 10^{-5}$ describes the thermal polarization of the system and I is a unit matrix, using the method of spatial averaging. From the state $\rho _{000}$, we prepared the input state $\rho _{in}=\frac{1-\epsilon }{8}\mathbf{1}+\epsilon |0\rangle _{1}|\Psi _{in}\rangle_{23}\langle \Psi _{in}|_{1}\langle 0|$ with a rotation and a CNOT gate. The probe qubit 1 was then put into a superposition state by another Hadamard gate. The physical evolutions were implemented by rotating system qubit and ancilla qubit in the magnetic fields $\mathbf{B_{s}}$ and $\mathbf{B_{a}}$, respectively, conditioned on the state of the probe qubit. We set $\omega_{s}t=\mu B_{s}t=2\pi $ so that the system qubit undergoes a cyclic evolution, i.e. $U_{s}$ becomes a unit operator. Therefore in the experiment, we only implemented a conditional rotation $U_{a}$ on the ancilla, where $U_{a}=U_{a}^{Sjo}$ and $U_{a}=U_{a}^{Uhl}$ for observing the Sj\"{o}qvist and Uhlmann phases, respectively.

In order to improve the quantum coherent control, experimentally every single qubit gates was created by using robust strongly modulating pulses (SMP)\cite{16,17,18}. We maximize the gate fidelity of the simulated propagator to the ideal gate, and we also maximize the effective gate fidelity by averaging over a weighted distribution of radio frequency (RF) field strengths, because the RF-control fields are inhomogeneous over the sample. Theoretically the gate fidelities we calculated for every pulse are greater than $0.995$, and the pulse lengths range from $200$ to $500$ $\mu s$. The quantum circuit of Fig. \ref{eps3} was realized with a sequence of these local SMPs separated by time intervals of free evolution under the Hamiltonian. The overall theoretical fidelity of this pulse sequence is about $0.98$.

The phase detection requires the measurement of the complex signal, $\langle \sigma _{-}^{p}\rangle $, containing the x- and y- components. In fact, in any NMR experiment, this complex signal $\langle \sigma _{-}^{p}\rangle (t)$ corresponds to the complex free induction decay (FID), obtained by simultaneous observation of both x- and y- components by quadrature detection. This signal was subjected to a complex Fourier transformation to obtain the complex spectrum, from which we extracted the real and imaginary spectra. The geometric phase $\gamma $ was achieved by integrating the multiplex resonance lines of the qubit. That is, $\gamma=arctan(Int_{Im}/Int_{Re})$ , where $Int_{Re}$ and $Int_{Im}$ represent, respectively, the integrals of the real and imaginary spectra.

\begin{figure}[tbph]
\centering
\includegraphics[width=1\columnwidth]{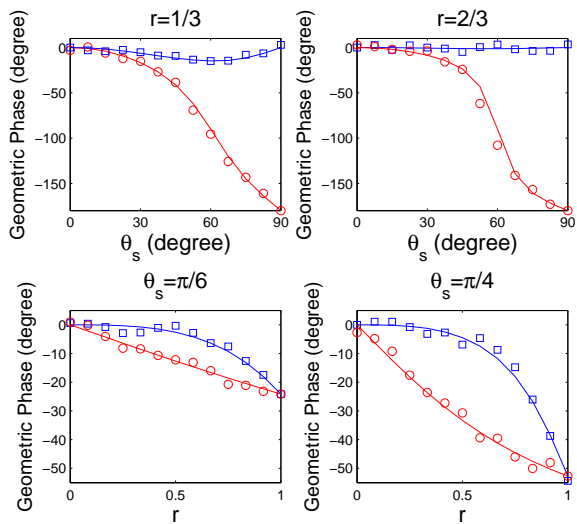}
\caption{(Color online) Experimental values for Uhlmann's geometric phases $\protect\gamma ^{U}$ (denoted by blue squares) and Sj\"{o}qvist's geometric phases $\protect\gamma ^{S}$ (denoted by red circles). Theoretical values are represented by smooth curves. Variability in the experimental points (estimated by repetition) was about ¡À3¡ã. The deviation of the measured data points from their theoretical values mainly resulted from the inhomogeneity of the radio frequency field and the static magnetic field, imperfect calibration of radio frequency pulses and unidentified systematic errors.}
\label{eps4}
\end{figure}
We measured both the Sj\"{o}qvist phase $\gamma ^{S}$ (using the unitary cyclic evolution of the ancilla $U_{a}^{Sjo}(r,\theta _{s})$) and the Uhlmann phase $\gamma ^{U}$ (using the unitary cyclic evolution of the ancilla $U_{a}^{Uhl}(r,\theta _{s})$) for mixed states of varying purity $r$ and varying angle of the magnetic field $\theta _{s}$. Two separate sets of experiments were performed: one is to vary the purity $r$ from 0 to 1 with a fixed the angle of the magnetic field $\theta _{s}$ (we chose two values of $\theta _{s}=\pi /6$ and $\theta _{s}=\pi /4$ in the experiments); the other
is to vary the $\theta _{s}$ value from 0 to $\pi /2$ with fixed purity $r$ (likewise, two values of $r=1/3$ and $2/3$ were chosen in the experiments). Data were collected for 13 equidistant values of the variable parameter, either the purity $r$ or the angle $\theta _{s}$.

Fig. \ref{eps4} shows the experimental results for these four situations. Clearly, the measured Sj\"{o}qvist and Uhlmann phases are in excellent agreement with the theoretical expectation (see Fig. \ref{eps2}). The deviation between the experimental and theoretical values is primarily due to the inhomogeneity of the radio frequency field and the static magnetic field, imperfect calibration of radio frequency pulses, and signal decay during the experiments.

In conclusion, we have experimentally observed the geometric phase of Uhlmann's phase and Sj\"{o}qvist's phase changing with the purity of mixed state. Also we find that different results of mixed state geometric phase correspond to different choice of the representation of Hilbert space of the ancilla. These are in accordance with the theoretical predictions. Our work raises a number of possible interesting future directions. Firstly, it seems that we can obtain other mixed state geometric phase definitions (to Uhlmann's and Sj\"{o}qvist's) by suitably tailoring our interaction with the environment. Secondly, it may be that some of these geometric phases are more robust than others, in which case, they are worth learning about. Thirdly, and more fundamentally, what happens it -- for some reason -- we cannot access the ancillary bit\cite{19,20,21}? Is the notion of mixed state geometric phase then to be abandoned, or -- more likely -- can we still use some reference with respect to which the geometric phase could be defined?

This work was supported by the National Natural Science Foundation of China, the CAS, Ministry of Education of PRC, and the National Fundamental Research Program. This work was also supported by European Commission under Contact No. 007065 (Marie Curie Fellowship). V.V. acknowledges the Wolfson Foundation and the Royal Society as well as the Engineering and Physical Science Research Council in UK. 

\end{document}